\documentclass[twoside,12pt,a4paper]{article}
\usepackage[dvips]{epsfig}
\usepackage{a4}
\usepackage{latexsym}
\usepackage{graphics}

\def\beq{\begin{equation}}
\def\eeq{\end{equation}}
\def\beqn{\begin{eqnarray}}
\def\eeqn{\end{eqnarray}}

\begin{document}
\title{Quark model predictions for the SU(6)-breaking ratio of the proton 
momentum distributions}
\author{M.M. Giannini, E. Santopinto, A. Vassallo\\
\small{
Dipartimento di Fisica dell'Universit\`a di Genova, I.N.F.N. 
Sezione di Genova, Italy}\\
\\
M. Vanderhaeghen\\
\small{Institut f\"{u}r Kernphysik, 
Johannes Gutenberg Universit\"{a}t, D-55099 Mainz, Germany}}

\date{\today}
\maketitle

\begin{abstract}

The ratio between the anomalous magnetic moments of proton and neutron 
has recently been suggested to be connected to 
the ratio of proton momentum fractions carried by valence quarks.
This ratio is evaluated using different constituent quark models, 
starting from the CQM density distributions and calculating the 
next-to leading order distributions.  We show that this 
momentum fraction ratio is a sensitive test 
for SU(6)-breaking effects and is a useful observable to distinguish 
among  different CQMs. We investigate also the possibility of getting 
constraints on the formulation of quark structure models.
 
PACS : 12.39.-x, 13.60.Hb, 14.20.Dh \\
Keywords: hadrons, partons, parton distributions, 
constituent quark models. 

\end{abstract}

\newpage

\section{Introduction}

The static properties of baryons are an important testing ground for
QCD based calculations in the confinement region. However, different 
CQMs\cite{is,ci,ferr,gi,olof,bil,plessas} are able to obtain 
a comparable good description of the low energy data, so that
it is difficult to discriminate among them. A fundamental aspect of the 
theoretical description is the introduction of terms in the quark Hamiltonian 
which violate the underlying $SU(6)-$symmetry. It is therefore important to 
find out observables which are sensitive to the various  SU(6)-breaking 
mechanisms.
    
In this respect, the relation proposed recently by Goeke, Polyakov and 
Vanderhaeghen \cite{gpv} between the 
anomalous magnetic moments of the proton and 
the neutron and the proton momentum fractions 
carried by valence quarks, $M_{2}^{q_{val}}$, 
might be a good candidate for testing SU(6)-breaking effects and 
can lead to important constraints on the models for the structure of the nucleon.


Quark models are able to reproduce in a extraordinary way the static 
low energy 
properties of  baryons with very few parameters and this gives us confidence 
that they are a good effective representation of the low energy strong interaction 
dynamics. The QCD based parton model reproduces in a beautiful way the $Q^2$ 
dependence of the high energy properties even with naive input. However the perturbative 
approach to QCD does not provide absolute values of the observables; one can only 
relate data at different momentum scales. The description based on the Operator Product 
Expansion (OPE) and the 
QCD evolution require the input of non-perturbative matrix elements which have to be 
predetermined \cite{roberts} and therefore the parton 
distributions are usually obtained in 
a phenomenological way from fits to deep inelastic lepton nucleon 
scattering and Drell-Yan processes. 
The basic steps are to find a parametrization \cite{martin-roberts} 
which is appropriate 
at a sufficiently large momentum ${{Q_0}^2}$, where it is expected that 
perturbation 
theory is applicable, and 
then QCD evolution techniques are used in order to obtain 
the parton distribution at higher $Q^2$. Using these parametrizations a 
large body of 
data is reasonably described, even if at the origin 
this parametrization is purely phenomenological.
  
Gluck, Reya and Vogt \cite{glueck-reya} started from a parametrized 
distribution of 
partons at a very low scale ${\mu^2}_{0}$, 
which resembles that of a naive Quark Model of hadron structure, 
in the sense that the contribution of the valence quarks to the 
structure function is dominant. 
As suggested by Parisi and Petronzio \cite{parisi}, 
the hadronic ${\mu^2_{0}}$ scale is defined
such that the fraction of the total momentum carried by the 
valence quarks is unity. 
This procedure opens the possibility of using Constituent Quark Models as input in 
order to calculate the nonperturbative (twist-two) nucleon  matrix elements, as proposed 
by Jaffe and Ross \cite{jaffe}.
 
The scheme developed by Traini et al.\cite{traini-vento} takes into account all these 
aspects: it uses as input the quark model results in order to determine the non 
perturbative matrix elements at the hadronic scale \cite{parisi}, then an upwards 
NLO evolution procedure at high momentum transfer ($Q^{2}=10$~GeV$^2$) is 
performed\cite{traini1}. 

Starting from three different Constituent Quark Models \cite{is,bil,ferr}, 
we have 
calculated the parton distributions at the hadronic scale and we have 
evaluated the ratio of the proton momentum fractions carried by valence
quarks. A NLO evolution has been performed up to $Q^2=10$~GeV$^2$.

All models give a good description of the spectrum and have been used also to 
describe various observables 
(elastic and inelastic form factors, strong decays). In particular, the 
different results for the electromagnetic transition form factors indicate 
that the models have a quite different $Q^2$-behaviour. 
However, as we shall see later, the ratio of the proton momentum
fractions carried by valence quarks  
is independent of the scale $Q^2$, therefore we expect that the study of this 
relation will give important information on general aspect of CQM.

The paper is organized as follow. In Section 2 we review in a critical 
way the new relation as found in Ref.~\cite{gpv} between the 
ratio of the anomalous magnetic moments of the proton and the neutron and the 
ratio of the proton 
momentum fractions $M_{2}^{q_{val}}$.  
In Section 3 the unpolarized parton distributions are evaluated, at the 
hadronic scale, using different CQMs, and an evolution procedure is 
performed and then in Sect. 4 the ratio of the proton momentum fractions
carried by valence quarks is calculated as a function 
of $Q^2$ and compared with experimental values and with the results of the
models for the ratio of the anomalous magnetic moments.

\section{Ratio of proton momentum fractions carried by valence quarks}

\begin{figure}[h]
\includegraphics[width=14cm]{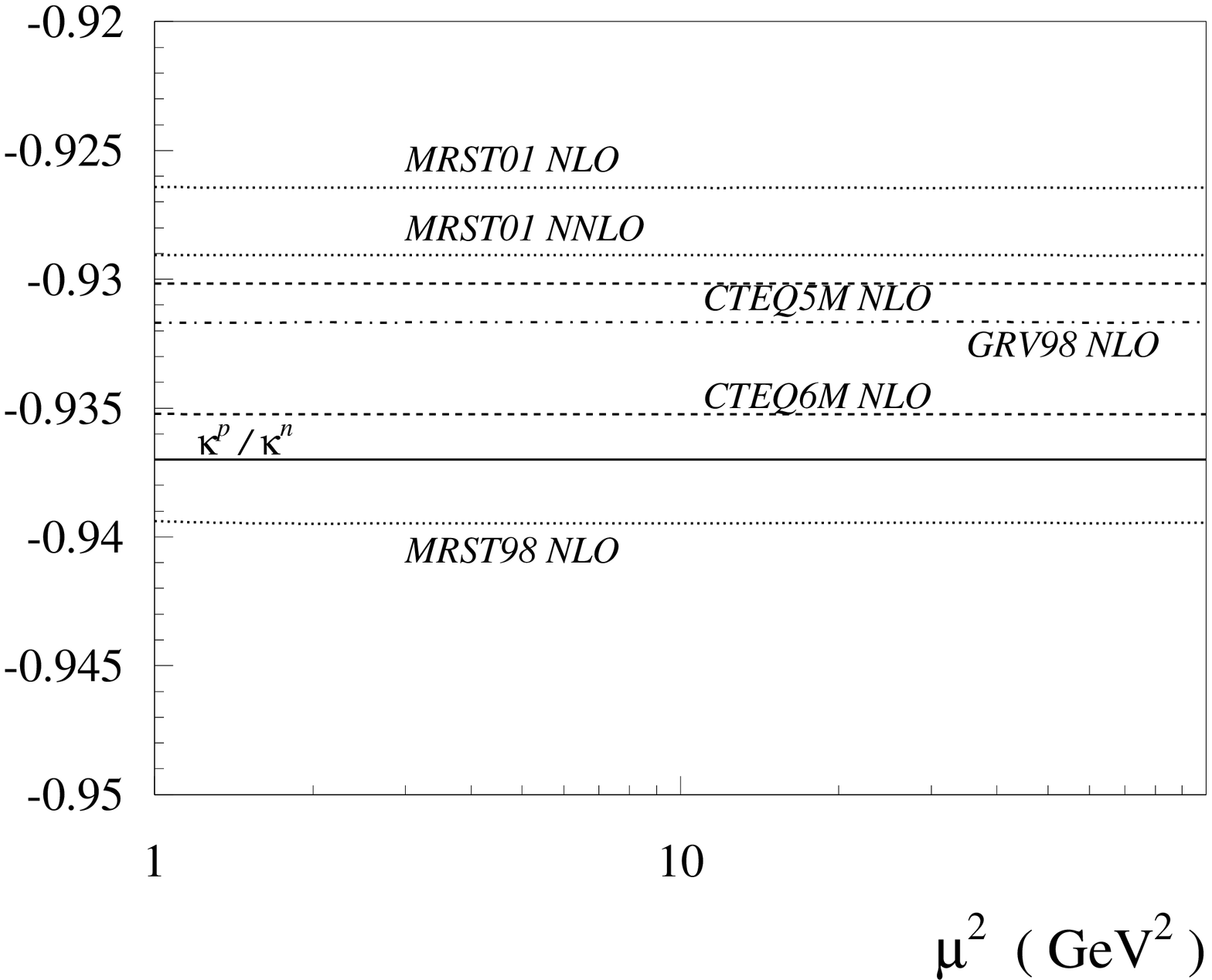}
\vspace{.75cm}
\caption[]{\small Scale dependence of the 
{\it rhs} of Eq.~(\ref{eq:kpkn}) for  
various phenomenological forward parton distributions as indicated on
the curves. 
Dotted curves : MRST parton distributions 
(MRST98 NLO \cite{Mar98}, MRST01 NLO \cite{Mar01}, MRST01 NNLO \cite{Mar02}). 
Dashed curves : CTEQ parton distributions  
(CTEQ5M NLO \cite{Lai00}, CTEQ6M NLO \cite{Pum02}). 
Dashed-dotted curve : GRV98 NLO(${\overline{\mathrm{MS}}}$)
\cite{Glu98}. Also shown is the {\it lhs} of Eq.~(\ref{eq:kpkn}), i.e. the 
experimental value for $\kappa^p / \kappa^n$ (constant solid curve).}
\label{fig:valrel}
\end{figure}

In Ref.~\cite{gpv}, a relation has been proposed between the ratio
of the proton and neutron anomalous magnetic moments and the momentum
fractions carried by valence $u$- and $d$-quark distributions, as follows~:
\begin{eqnarray}
\frac{\kappa^p}{\kappa^n} \,=\, -\,{1 \over 2} \, 
\frac{4 \, M_2^{d_{val}}+ M_2^{u_{val}}}{ M_2^{d_{val}}+ M_2^{u_{val}}}\, ,
\label{eq:kpkn}
\end{eqnarray}
with the proton momentum fraction carried by the valence quarks defined as 
\begin{equation}
M_2^{q_{val}} \,=\, \int_0^1dx \, x \, q_{val}(x) \, .
\label{mevalproton}
\end{equation}\\
In Fig.~\ref{fig:valrel}, we show the  
scale dependence of the {\it rhs} of Eq.~(\ref{eq:kpkn}), which we shall
henceforth denote with R,   
for various recent parametrizations of 
next-to-leading order (NLO) and 
next-to-next-to-leading order (NNLO) parton distributions.
Fig.~\ref{fig:valrel} shows 
that the scale dependence drops out of the {\it rhs} of Eq.~(\ref{eq:kpkn}),
although the numerator and denominator separately clearly have a scale
dependence. Furthermore, it is seen from Fig.~\ref{fig:valrel}, 
for all NLO and one NNLO parametrizations of
parton distributions, that the relation of Eq.~(\ref{eq:kpkn}) is
numerically verified to an accuracy at the one percent level! 
In particular, the most recent MRST01 NLO \cite{Mar01}, the  
MRST01 NNLO \cite{Mar02}, and the CTEQ6M NLO \cite{Pum02} parton 
 distributions (which appeared after the writing of Ref.~\cite{gpv}),
nicely confirm the finding of Ref.~\cite{gpv}. 
Although the relation Eq.~(\ref{eq:kpkn}) was originally derived
within a parametrization of generalized parton distributions, it is in
fact completely independent of such a parametrization, as the 
{\it rhs} of  Eq.~(\ref{eq:kpkn}) is expressed in terms of moments of
forward valence quark distributions alone. 
\newline
\indent
The above observations from phenomenology 
suggest that Eq.~(\ref{eq:kpkn}) holds and that 
the unpolarized valence $u-$ and $d$-quark forward 
distributions contain a non-trivial information 
about the anomalous magnetic moments of the proton and neutron. 
It is the aim of the present work to investigate the relation 
of Eq.~(\ref{eq:kpkn}) in different quark models. 
\newline
\indent
Let us firstly consider the simplest quark model, 
with exact $SU(6)$ symmetry. In this limit, 
$M_2^{u_{val}} = 2 \,M_2^{d_{val}}$, and 
$\kappa^p$ = - $\kappa^n = 2$, so that one 
immediately verifies that Eq.~(\ref{eq:kpkn}) holds. 
\newline
\indent
In reality, the ratio of anomalous magnetic moments deviates from the 
$SU(6)$ limit by about 6.5 \%. The smallness of this 
deviation is the main reason why constituent quark models 
are quite successful in predicting 
nucleon (and more generally baryon octet) magnetic moments. 
In quark model language, 
the relation of Eq.~(\ref{eq:kpkn}) implies that the small breaking of the 
$SU(6)$ symmetry follows some rule which is encoded in the valence quark
distributions. In particular, it is interesting to investigate 
a possible correlation between the ratio of valence $d-$ and $u$-quark
distributions, and the ratio of proton to
neutron anomalous magnetic moments in different models. 
To this end, we turn in the next
section to the calculation of parton distributions in quark models
with different $SU(6)$ breaking mechanisms. 

\section{Parton distributions from quark models}

The approach, recently developed by M. Traini et al. for the 
unpolarized distributions  \cite{traini-vento}, connects the model wave 
functions and the parton distributions at the input hadronic scale 
through the quark momentum density distribution.
In the unpolarized case one can write the parton 
distributions \cite{traini-vento}:
\begin{equation}
q_V(x,\mu_0^2)=\frac{1}{(1-x)^2}~\int d^3 k~ n_q(|\mbox{\bf k}|) ~
\delta(\frac{x}{1-x}-\frac{k_+}{M})
\label{partdistr1}
\end{equation}
where $k_+$ is the light-cone momentum of the struck parton, 
and $n_q(|\mbox{\bf k}|)$
represents the density momentum distribution of the valence quark of q-flavour:
\begin{eqnarray}
n_u(|\mbox{\bf k}|)=\langle N,J_z=+1/2|\sum_{i=1}^3 \frac{1+\tau_i^z}{2}
     ~\delta(\mbox{\bf k}-\mbox{\bf k}_i)|N,J_z=+1/2\rangle \nonumber \\
n_d(|\mbox{\bf k}|)=\langle N,J_z=+1/2|\sum_{i=1}^3 \frac{1-\tau_i^z}{2}
~\delta(\mbox{\bf k}-\mbox{\bf k}_i)|N,J_z=+1/2\rangle~,
\label{momdistr}
\end{eqnarray}
\noindent $\tau_i^z$ is the third component
of the isospin Pauli matrices, $k_{i}$
is the momentum of the $i$th constituent quark in the CM frame of the nucleon,
$|N,J_z=+1/2\rangle $ is the 
nucleon wave function (in momentum space) with $J_z=+1/2$ component.\\
Using $k_+=k_0+k_z$, one can integrate eq. \ref{partdistr1} over 
the angular variables and get:
\begin{equation}
q_V(x,\mu_0^2)
=\frac{2\pi M}{(1-x)^2}\int_{k_m(x)}^\infty d |\mbox{\bf k}|
|\mbox{\bf k}| ~n_q(|\mbox{\bf k}|)\mbox{,}
\label{partdistr}
\end{equation}

\noindent where
\begin{displaymath}
k_m(x)=\frac{M}{2}~ \Big|~\frac{x}{1-x}-\left (\frac{m_q}{M}\right )^2~
\frac{1-x}{x}~\Big|~,
\end{displaymath}
$M$ and $m_{q}$
are the nucleon and (constituent) quark masses respectively.\\
Eq.~(\ref{partdistr}) can be applied to a large class of quark models and 
satisfies some important requirements:
it vanishes outside the support region $0 \le x \le 1$ and it has the 
correct integral property in order to preserve the number normalization.

In the present section we shortly illustrate the evolution procedure we have 
been using. Even if alternative factorization schemes have been investigated, 
we remain within the $\overline {MS}$
renormalization and 
DIS factorization scheme(see \cite{traini1} and references therein). In this case the moments 
of the $F_2$ proton (neutron) structure functions have the simple expression
\begin{eqnarray}
\langle F_2^{\rm p,(n)}(Q^2)\rangle_n & = &
\sum_{q=u,d,s}\,e^2_q \langle x\,q(x,Q^2) + x\,\bar q(x,Q^2)\rangle_n =
\nonumber \\
& = & 2\,\left[+ (-) {1\over12}\,\langle x q_3(Q^2)\rangle_n + 
{1\over36}\,\langle  xq_8(Q^2)\rangle_n + {1\over9}\,
\langle x \Sigma(Q^2)\rangle_n \right] 
\,\, ,
\label{ev:25}
\end{eqnarray}
where $+\,(-)\,$ refers to proton and neutron respectively;
$ \Sigma = \sum_q ( q +  \bar q)$ is a singlet component and 
$ q_3 =  u +  \bar u -( d +  \bar d)$, $ q_8 =  u +  \bar u +  
d +  \bar d - 2\, ( s +  \bar s)$ are nonsinglet (NS) contributions.
The Wilson coefficients $ C^{(1),q}_n$ and $ C^{(1),g}_n\,$,
in the $\overline {MS}$ renormalization and factorization
scheme, can be found,e.g., in Refs.\cite{GRV92,WM96}.

The NLO evolution of the unpolarized distributions is performed following the
solution of the renormalization group equation in terms of moments, i.e.
$\langle f(Q^2)\rangle_n = \int_0^1 dx\,f(x,Q^2)\,x^{n-1}$.
Since, in our case, the starting point for the evolution ($\mu_0^2$) is rather low, 
the form of the equations must guarantee complete symmetry for the evolution 
from $\mu_0^2$ to $Q^2 \gg \mu_0^2$ and {\em back~}
avoiding additional approximations associated with Taylor expansions and not
with the genuine perturbative QCD expansion \cite{traini1}. In particular for 
the Non-Singlet sector we have
\begin{equation}
\langle  q_{NS}(Q^2)\rangle_n = \left[\left( \frac{\alpha(Q^2)}
{\alpha(\mu_0^2)}\right)^{\frac{\gamma_{NS}^{0,n}}{2\beta_0}}
\frac{1+\left( \frac{\gamma_{NS}^{1,n}}{2\beta_0}-
\frac{\gamma_{NS}^{0,n}\beta_1}{2\beta_0^2}
\right) \frac{\alpha(Q^2)}{4\pi}}{1+\left(\frac{\gamma_{NS}^{1,n}}{2\beta_0}-
\frac{\gamma_{NS}^{0,n}\beta_1}{2\beta_0^2}
\right) \frac{\alpha(\mu_0^2)}{4\pi}}\right]
\langle  q_{NS}(\mu_0^2)\rangle_n\,\, ,
\label{ev:18}
\end{equation}
where $ \gamma_{NS}^{(0,1),n}$ are the anomalous dimensions at LO and NLO
in the DIS scheme \footnote{The $\gamma_{NS}^{1,n}$ are redefined in the DIS scheme 
in such a way that the Eq.~(\ref{ev:25}) holds, i.e.
$\gamma_{NS}^{1,n} \to \gamma_{NS}^{1,n} + 2\,\beta_0\,C^{(1),NS}_{n}$
.}, 
and $\beta_0$,
$\beta_1$ the expansion coefficients (up to NLO) of the function $\beta(Q^2)$:
$\beta_0 = 11 - 2/3\,N_f$, $\beta_1 = 102 - 38/3\,N_f$ for $N_f$ active flavors.
Eq.~(\ref{ev:18}) reduces to the more familiar form (e.g. Ref.\cite{glueck-reya,WM96})
\begin{equation}
\langle  q_{NS}(Q^2)\rangle_n = \left[\left( \frac{\alpha(Q^2)}{\alpha(\mu_0^2)}
\right)^{\frac{\gamma_{NS}^{0,n}}{2\beta_0}}
\left(1+\left(\frac{\gamma_{NS}^{1,n}}{2\beta_0}-\frac{\gamma_{NS}^{0,n}\beta_1}
{2\beta_0^2}
\right) \left(\frac{\alpha(Q^2)-\alpha(\mu_0^2)} {4\pi}\right)\right)\right]\,
\langle  q_{NS}(\mu_0^2)\rangle_n
\label{ev:19}
\end{equation}
after performing a Taylor expansion for {\em both~} 
${\alpha(\mu_0^2)\over 4\pi} \ll 1$ and 
${\alpha(Q^2)\over 4\pi} \ll 1$.

The $\Lambda$'s values are suggested by the analysis of 
Gl\"uck {\em et al.}\cite{glueck-reya}, $\left.\alpha_s(\mu_0^2)\right|_{\rm NLO}$ 
is obtained evolving back the
valence distribution as previously mentioned, and $\mu_0^2$ is found 
by solving numerically the NLO transcendental equation 
\begin{equation}
\ln {\mu_0^2\over\Lambda_{\rm NLO}^2}-{4\,\pi\over\beta_0\,\alpha_s} + 
{\beta_1\over\beta_0^2}\,\ln\left[
{4\,\pi\over\beta_0\,\alpha_s} + {\beta_1\over\beta_0^2}\right] = 0\,,
\label{0:10}
\end{equation}
which assumes the more familiar expression
\begin{equation}
{\alpha_s(Q^2) \over 4\pi}={1 \over \beta_0\ln(Q^2/\Lambda_{\rm NLO}^2)}
\left(1-{\beta_0 \over \beta_0^2}\,{\ln\ln(Q^2/\Lambda_{\rm NLO}^2)\over
\ln(Q^2/\Lambda_{\rm NLO}^2)}\right)
\label{1:10}
\end{equation}
only in the limit $Q^2\gg\Lambda_{\rm NLO}^2$;
(an interesting discussion on the effects of the approximation (\ref{1:10}) 
can be found in Ref.\cite{WM96}).

The actual value of $\mu_0^2$ is fixed
evolving back (at the appropriate perturbative order) unpolarized
data fits, until the valence distribution 
$x\,V(x,\mu_0^2) = x\,u_V(x,\mu_0^2)+x\,d_V(x,\mu_0^2)$ matches the required 
momentum ($\int dx\,x\,V(x,\mu_0^2) = 1$). The resulting NLO (LO)
parameters are \cite{traini1}:
\begin{eqnarray}
\,\,\, && \left.{\alpha_s(\mu_0^2) \over 4\,\pi}\right|_{\rm NLO} = 0.142\,, 
\,\,\,\, \left.\mu_0^2\right|_{\rm NLO} = 0.094\, {\rm GeV}^2\,,\,\,\,\, 
\Lambda_{\rm NLO} = 248\, {\rm MeV} \,;\nonumber \\
&& \left.{\alpha_s(\mu_0^2) \over 4\,\pi}\right|_{\rm LO}\,\,\, = 0.290\,,
\,\,\,\, \left.\mu_0^2\right|_{\rm LO}\,\,\, = 0.079\, {\rm GeV}^2\,,\,\,\,\, 
\Lambda_{\rm LO}\,\,\, = 232\, {\rm MeV}\,.
\label{0:9}
\end{eqnarray}\\

We discuss the results obtained using different models for the 
valence quark contributions, namely the Isgur-Karl (IK) 
model \cite{is}, which has been largely used in the past to study the 
low-energy properties of hadrons and also deep 
inelastic polarized and unpolarized scattering\cite{traini1}, a 
hypercentral Coulomb-like plus linear confinement 
potential model \cite{ferr} inspired by lattice QCD \cite{lat}
and an algebraic model \cite{bil}; 
the wave functions of the last two models give a rather good 
description of the electromagnetic elastic and transition form factors 
\cite {gi} \cite{mds} \cite{bil} \cite{iacem}.\\

1) The well known Isgur Karl model is based on a harmonic oscillator potential 
plus a One-Gluon-Exchange-hyperfine interaction which is responsible 
for the $SU(6)$ breaking of the symmetry.
The nucleon wave function is written as
a superposition of $SU(6)$ configurations, that is 
\beq
|N \rangle~=~a_S |56, 0^+ \rangle~+~a_S' |56', 0^+ \rangle~+~a_M |70, 0^+ \rangle~+~a_D |70,2^+ \rangle~ .
\eeq
In particular we discuss the result for the Isgur Karl model(IK) $a_S=0.931, 
a_S'=-0.274,
a_M=-0.233, a_D=-0.067$ and also for a simplified model where only the $a_S$ and 
$a_M$ (or $a_D$) coefficients
do not vanish. The contributions from the SU(6) breaking components come from
the amplitudes $a_S'$, $a_M$ and $a_D$ of the 
$|56', 0^+ \rangle$  $|70, 0^+ \rangle$ 
and $|70,2+ \rangle$ multiplets, since without the OGE-hyperfine interaction 
 $a_S'=a_M=a_D=0$.

The corresponding momentum density distributions are 

\beqn
n_u(|\mbox{\bf k}|)& = & {1\over 2}{1\over \alpha^3 \pi^{3/2}}
\left({3\over 2}\right)^{3/2}\left\{ 4 \left[a_S^2 + {a_S'}^2 \left({5\over 4} - {3\over 2}{{\bf k}^2 \over \alpha^2} + {3\over 4}{{\bf k}^4 \over \alpha^4}\right) + a_M^2 \left({5\over 8} - {1\over 4}{{\bf k}^2 \over \alpha^2} + {3\over 8}{{\bf k}^4 \over \alpha^4}\right) \right]\right.\nonumber \\
& + & a_D^2 \left({1\over 2} + 3 {{\bf k}^2 \over \alpha^2} + {3\over 10}{{\bf k}^4 \over \alpha^4}\right) -a_S a_S' 4 \sqrt{3} \left(
{{\bf k}^2 \over \alpha^2} - 1 \right) + a_S a_M \sqrt{6} \left(
{{\bf k}^2 \over \alpha^2} - 1 \right) \nonumber \\
& + & \left. a_S' a_M \sqrt{2} \left(-{1\over 2} + 3 {{\bf k}^2 \over \alpha^2} - {3\over 2}{{\bf k}^4 \over \alpha^4}\right) \right\}
\, e^{-{3\over 2}{{\bf k}^2 \over \alpha^2}}\label{eq:nukik}
\eeqn
\beqn
n_d(|\mbox{\bf k}|)& = & {1\over 2}{1\over \alpha^3 \pi^{3/2}}\left({3\over 2}\right)^{3/2}\left\{2 \left[a_S^2 + {a_S'}^2 \left({5\over 4} - {3\over 2}{{\bf k}^2 \over \alpha^2} + {3\over 4}{{\bf k}^4 \over \alpha^4}\right) + a_M^2 \left({5\over 8} - {1\over 4}{{\bf k}^2 \over \alpha^2} + {3\over 8}{{\bf k}^4 \over \alpha^4}\right) \right] + \right. \nonumber \\
& + & a_D^2 \left(1 + {3\over 5}{{\bf k}^4 \over \alpha^4}\right) 
-a_S a_S' 2 \sqrt{3} \left({{\bf k}^2 \over \alpha^2} - 1 \right) 
- a_S a_M \sqrt{6} \left({{\bf k}^2 \over \alpha^2} - 1 \right) + \nonumber \\
& - & \left. a_S' a_M \sqrt{2} \left(-{1\over 2} + 3 {{\bf k}^2 \over \alpha^2} - {3\over 2}{{\bf k}^4 \over \alpha^4}\right) \right\}
\, e^{-{3\over 2}{{\bf k}^2 \over \alpha^2}}\label{eq:ndkik}
\eeqn
with
$$
\int n_u(|\mbox{\bf k}|) d\mbox{\bf k} = 2 \int n_d(|\mbox{\bf k}|) d\mbox{\bf k} = 2 
$$
and 
$$
n(|\mbox{\bf k}|) = n_u(|\mbox{\bf k}|) + n_d(|\mbox{\bf k}|)~.
$$

The ensuing $M_2$ momenta for the u and d quarks in the proton are reported in 
Fig. 2. The scale dependence for the single momenta is quite smooth apart
from the low $Q^2$-values.\\
  
2) The hypercentral Constituent Quark Model (hCQM) is based on 
a Coulomb-like potential plus a linear confining potential to which a 
OGE-hyperfine interaction is added. 
The difference with the IK model is mainly in the spatial wave functions which 
are not gaussians, but are more spread out 
and are obtained by numerical solution 
of the $3-$quark wave equation. Moreover, the nucleon state is written as a 
superposition of five SU(6)- configurations:
\beq
|N \rangle~=~a_S |56, 0^+ \rangle~+~a_S' |56', 0^+ \rangle~+~a_S'' | 56'', 0^+
\rangle ~+~a_M |70, 0^+ \rangle~+~a_D |70,2^+ \rangle~
\eeq
with 
$a_S=0.9997, a_S'=0.0217,a_S''=0.0041,a_M=0.0038, a_D=-0.0012$.
Without the OGE-SU(6) breaking term  $a_S'=a_S''=a_M=a_D=0$.
The resulting momentum density distribution contains higher 
momentum components in comparison with the h.o one.\\

2\_bis) The SU(6) invariant hamiltonian is left unchanged, while the 
SU(6)-breaking mechanism is provided by a spin- and isospin-dependent 
interaction \cite{hcqmiso}. Here also the content of high momentum component 
is greater than in the h.o. case.\\

3) In the model proposed by Iachello et al.\cite{bil} 
the hamiltonian consists of a part corresponding to 
the vibration and rotation of a top to which a G\"ursey-Radicati  
spin and isospin dependent term is added. The G\"ursey-Radicati term is 
diagonal with respect to the SU(6)-configurations, so it splits but does not mix the $SU(6)-$configurations.
An $SU(6)$-breaking mechanism is implemented in a phenomenological way considering  different $u$ and $d$ charge distributions \cite{iacem}, which correspond to different $u$ and $d$ effective charge radii. 
In this way the nucleon elastic form factors are obtained folding the top form factors with the $u$ and $d$ charge distributions (assumed to be exponential-like) and the results have a dipole behaviour.

Also in this case the momentum density distribution 
contains high momentum components and one can imagine that this 
will strongly influence the results.

The unpolarized valence quark distributions are given by \cite{privatecom}
\beq
n_q(|\mbox{\bf k}|)= ~ \frac{8}{\pi^2~a_q^3}\frac{{\mathcal{N}}_q}{\left( 1 + 
\frac{\mbox{\bf k}^2}{a_q^2} \right)^4}
\eeq
where:
$$
{\mathcal{N}}_u=2 \hspace{2cm} {\mathcal{N}}_d=1 $$
$$
a_u^{-1}=0.258~ \mbox{fm} \hspace{2cm} a_d^{-1}=0.285~\mbox{fm}.$$\\

The validity of Eq.~(\ref{eq:kpkn}) for the model 2\_bis
is analyzed in Fig. 3. The two members are equal within 0.2 \%, although the 
$\kappa$-ratio differs by about 7 \% from the experimental value 
($\sim -0.937$).

Similar results, reported in Table I, hold for the other models, with the 
exception of the U(7) model, where the $\kappa$-value is correctly reproduced 
by construction, while the equation is violated up to a few percent.

In order to test if this feature depends on the choice of the CQMs or
 is a 
general characteristic, we have used the analytic expression supplied by the 
Isgur-Karl model and tried to reproduce the experimental value of the two 
ratios by leaving the amplitudes $a_S'$,$a_M$ and $a_D$ free. One can also
vary the h.o. constant $\alpha$, with $\alpha^{-1}$ being a measure of the
confinement radius. The $Q^2$-behaviour of the I.K. model is unrealistic
because of the gauss-factors, however also in this case the ratio is quite
scale
independent. The procedure of fitting the amplitudes corresponds to
introduce implicitly quite different hamiltonians.
The anomalous magnetic moments have the following expressions: 
\begin{eqnarray}
\kappa_p &= & 2(1-a_M^2)-4a_D^2 \nonumber \\
\kappa_n &= & -2(1-a_M^2)+3/2 ~a_D^2~.        
\label{eq:momanom}
\end{eqnarray}
If one adopts a model where the only SU(6) breaking comes from the $a_M$, 
it is immediately seen from equation (\ref{eq:momanom}) that the 
$\kappa$-ratio is exactely equal to -1, like in the SU(6) limit.
The crucial quantity seems then to be the $a_D$ amplitude. Assuming that 
the D-wave amplitude is the only SU(6)-breaking term 
({\em D-model}), we have that:
$$\frac{2 a_S^2 - a_D^2}{-2a_S^2 - 1/2~a_D^2}=-0.937$$
if $a_S=0.955$ and $a_D=0.295$. 
Calculating the {\it rhs} of Eq.~(\ref{eq:kpkn}), which we refer as R in the following, 
 with these 
 two values of the 
parameter and varying $\alpha$ in a quite large interval, the best value 
obtainable is $R=0.9988$, with $\alpha=2.1~fm^{-1}$, differing by about 7\% 
from the $\kappa$-ratio.  
Finally, leaving completely free the amplitudes $a_S'$, $a_M$ and $a_D$ in 
order to fit the $\kappa$-ratio and R separately, the resulting amplitudes 
turn out to be complex.

Therefore, the proposed Equation (\ref{eq:kpkn}) seems to be valid (up to few 
percent) for all Constituent Quark Models provided that the SU(6)-violation is 
not too strong, but both values are quite far from the experimental value 
of the $\kappa$-ratio of $-0.937$. If one tries to force the SU(6)-violation 
to reproduce the experimental value, one is apparently faced with too strong 
constraints coming from the CQM itself. This is a possible indication that the 
degrees of freedom introduced in the current CQM may be inadequate since one 
has to take into account pion cloud effects \cite{Dahiya,Cloet}.   

\section{Discussion and Conclusions}
The relation Eq.~(\ref{eq:kpkn}) between the ratio of the proton and neutron anomalous 
magnetic moments and the momentum fractions 
carried by valence quarks, $M_{2}^{q_{val}}$,  
is exactly verified in the SU(6)-invariant limit, where both are equal to 
-1.

In the currently used Constituent Quark Models, SU(6) violations are 
introduced in different ways (One-Gluon-Exchange interaction, spin and/or 
isospin dependent terms, G\"ursey-Radicati mass formula, One-Boson-Exchange 
...). Such SU(6) violation is necessary in order to bring the anomalous proton
and neutron magnetic moments closer to the experimental values or 
to reproduce 
important features of the spectrum, such as the N-$\Delta$ mass difference.\\
In all the models we have considered in this paper (see Table I) the equality 
of Eq.(\ref{eq:kpkn}) holds within a few percent accuracy. This agreement is 
based on what all the CQMs have in common: the effective degrees of freedom of 
the three constituent quarks and the underlying SU(6) symmetry.\\
On the other hand, the experimental value of the ratio is not reproduced by 
CQMs, at variance with the calculations based on
 phenomenological parton distributions reported in Fig. 1. This means that 
the SU(6)-breaking mechanism contained in the phenomenological partonic 
distributions does not correspond to the SU(6) breaking mechanism implemented 
in the CQMs we have analyzed.\\
The quark densities as given in Eqs. (\ref{eq:nukik},\ref{eq:ndkik}) are 
evaluated in the 
rest frame, as we are using non-relativistic wavefunctions in this paper.
It is clear that in this way, relativistic boost effects are not included. 
Further work to quantify these relativistic boost effects is underway, even if
 we do not expect them to change in any important way our 
conclusion for the ratio of Eq. (\ref{eq:kpkn}).\\ 
To conclude, it seems that all CQMs are too strongly constrained by the 
presence of the standard degrees of freedom corresponding to three constituent quarks. 
Therefore additional degrees of freedom should be introduced, in particular quark antiquark pairs and/or 
gluons and the discussed equation 
of Ref.~\cite{gpv}, being sensitive 
to the SU(6)-breaking mechanism, will provide a useful tool for testing the 
new models.         

\section*{Acknowledgments}
The authors are indebted with Prof. Marco Traini for help and discussions 
concerning the evolution equation.

\clearpage
\newpage

\begin{figure}[h]
\includegraphics[width=14cm]{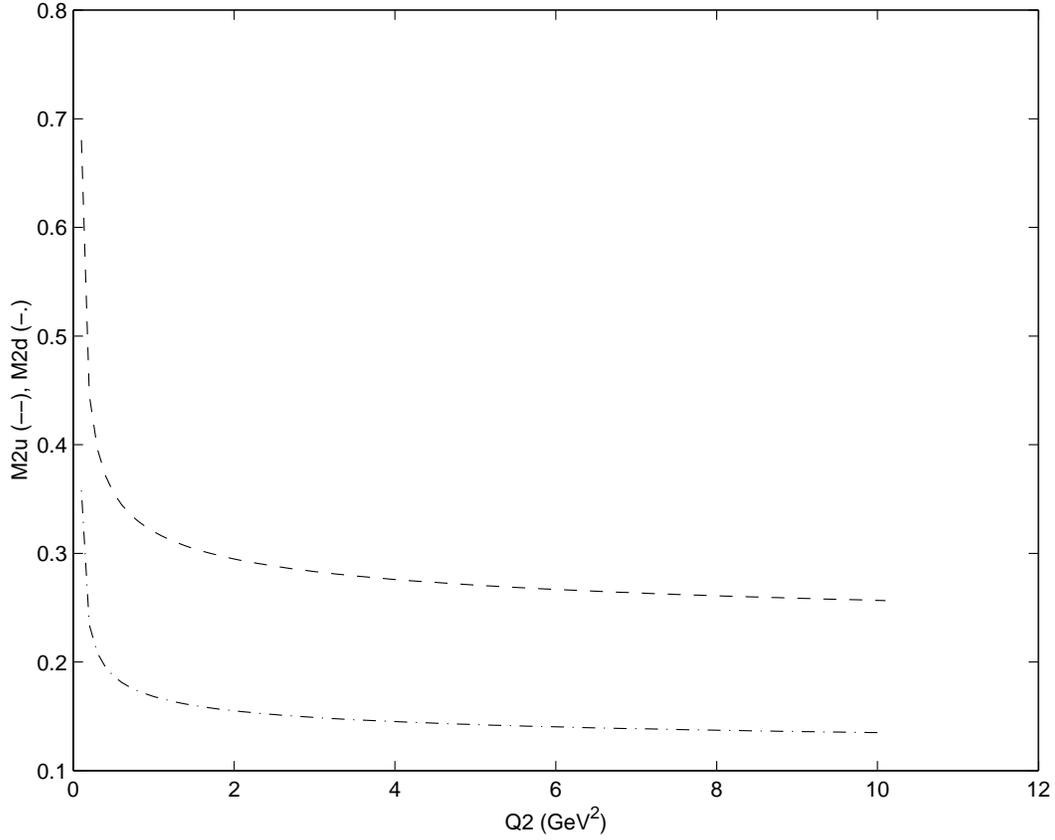}
\vspace{.75cm}
\caption[]{\small Scale dependence of the proton momentum fraction 
$M_2^{q_{val}}$ calculated with the Isgur-Karl model.}
\label{fig:moments}
\end{figure}

\begin{figure}[t]
\hspace*{-1cm}
\begin{picture}(480,390)
        \put(0,0){\makebox(480,390){
        \includegraphics[width=14cm]{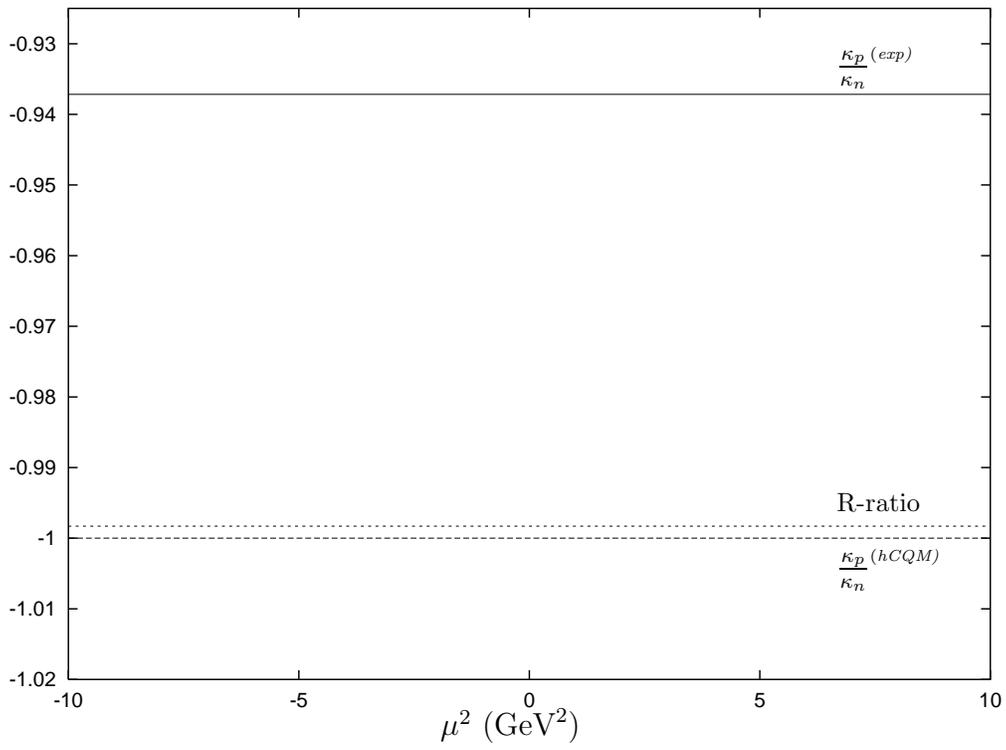}}}
        \put(220,50){$\mu^2~(\mbox{GeV}^2)$}
        \put(370,300){$\frac{\kappa_p}{\kappa_n}^{\mbox{\tiny (\em exp)}}$}
        \put(370,110){$\frac{\kappa_p}{\kappa_n}^{\mbox{\tiny 
              (\em hCQM)}}$}
        \put(370,135){\footnotesize $\mbox{R-ratio}$}
\end{picture}
\caption[]{The R-ratio for the HCQM  with isospin dependence compared
with
the $\kappa$-ratio calculated with the same
model and with the  
experimental value of the $\kappa$-ratio . }
\label{fig:momentumdensitydistr}
\end{figure}

\begin{table}[t]
\begin{tabular}{|c|c|c|c|c|}
\hline 
 &       &             &                 &    \\
 & I.K.  & HCQM + OGE  & HCQM + Isospin  & U7 \\
 &          &            &                 &                           \\
\hline 
 &          &            &                 & \\
Model prediction for $\frac{\kappa_p}{\kappa_n}$  & -1.0 & -1.0 & 
-1.0 & -0.9372\\
 & & & & \\
R-ratio at $Q^2=0.5 ~\mbox{GeV}^2$  & -1.0098 & -1.0030 & -0.9983 & -0.9881 \\
R-ratio at $Q^2=5.0 ~\mbox{GeV}^2$  & -1.0098 & -1.0030 & -0.9983 & -0.9881 \\
R-ratio at $Q^2=10.0 ~\mbox{GeV}^2$  & -1.0098 & -1.0030 & 
-0.9983 & -0.9881 \\
 & & & & \\
\hline 
\end{tabular}
\vspace{.75cm}

\caption[]{Different CQM predictions for the R-ratio and for the
$\kappa$-ratio $~\kappa^p / \kappa^n$ }
\end{table}

\end{document}